\documentstyle[12pt]{article}
\author{}
\date{}

\newcommand\gir[1]{\parbox[t]{1.2em}{#1

\vspace*{-\baselineskip}
\hspace*{0.17mm}#1

\vspace*{-\baselineskip}
\hspace*{0.34mm}#1}\!\!\!\!}

\begin{document}
\thispagestyle{empty}
\bibliographystyle{plain}
\begin{center}

\null
\vskip-1truecm
\rightline{IC/96/116}
\vskip1truecm
United Nations Educational Scientific and Cultural Organization\\
and\\
International Atomic Energy Agency\\
\medskip
INTERNATIONAL CENTRE FOR THEORETICAL PHYSICS\\
\vskip1.5truecm
{\bf COLLAPSE VERSUS TURBULENCE\footnote{\normalsize Submitted to
Physical Review E (Rapid Communications).}}\\
\vspace{1.5cm} 
L.A. Abramyan\\
 Institute of Applied Physics, Russian Academy of 
Sciences,\\ 
46 Ul'yanov Street, 603000 Nizhny Novgorod, Russian Federation,\\
\bigskip
\medskip
V.I. Berezhiani\quad and\quad  A.P. Protogenov\footnote{\normalsize
Permanent address:
 Institute of Applied Physics, Russian Academy of 
Sciences,
46 Ul'yanov Street, 603000 Nizhny Novgorod, Russian Federation.}\\
International Centre for Theoretical Physics, Trieste, Italy.\\ 
\end{center}
\vspace{1cm}
\centerline{ABSTRACT}
\baselineskip=18pt
\bigskip

We study the solutions of the equations 
of motion in the gauged (2+1)-dimensional nonlinear Schr\"odinger 
model. The contribution of Chern-Simons gauge fields 
leads to a significant decrease of the critical power 
of self-focusing. We also show that at appropriate 
boundary conditions in the considered model there exists 
a regime of turbulent motions of hydrodynamic type. 
\vspace{0.5cm}

\baselineskip=14pt
\noindent
\vspace{2cm}
\begin{center}
MIRAMARE -- TRIESTE\\
\medskip
July 1996
\end{center}
 

\newpage 

\baselineskip=18pt

\subsubsection*{1. Introduction}

The nonlinear Schr\"odinger equation (NSE) is one of the basic 
models for nonlinear waves. The conventional 
field for NSE application is nonlinear optics \cite{Tal,Vla} 
wherein it describes propagation of wave beams in dispersive   
nonlinear media. The NSE appears also when one considers various 
nonlinear waves in hydrodynamics and plasma physics \cite{Lit}. 
A most important application area is the problem of detailed 
description of collapsing field distributions for 
the NSE with local cubic nonlinearity \cite{Lms,Ras}. 
In the case of the opposite sign of nonlinearity the NSE is used as 
the basic model \cite{Kuz} of the low-dimension field theory  
to describe vortices in the problem of Bose condensation \cite{Gin}.  

Recent interest in problems with NSE solving in spatially $2D$ 
systems was associated with the account of the specific 
character of $2+1 D$ systems, which is manifested in equipping 
the NSE with a gauge field by replacing the usual derivatives by  
covariant ones. By that, the gauge field satisfies its 
own equation of motion (EM) with current, determined by the NSE 
solution. In the infrared limit, the main contribution to the 
EM of the gauge field in $2+1 D$ system is given by the 
Chern-Simons (CS) term within the system under consideration. 
The contribution of the gauge fields compensates in the Hamiltonian 
the contribution of nonlinearity at a certain relation of coupling 
constants. It leads to a soliton distribution of the fields  
found in Ref. \cite{Jac}. The nature of this phenomenon can easily 
be understood if one takes into 
account that CS term breaks ${\cal T}$\/- and ${\cal 
P}$\/-inversion symmetry of time and space. The chosen  
direction of the vector in the direction perpendicular to the 
plane can be considered as a chosen direction of rotation in 
the plane leading to the appearance of efficient repulsion.  
When this repulsion compensates the attraction, the Hamiltonian 
turns out to be limited at the bottom, and the CS solitons \cite{Jac} 
correspond to its zero value. These field distributions are solutions 
of the  self-dual equations \cite{Bog}. 

The results of the Refs. \cite{Jac,Hon,Wei} stimulated a 
series of papers in this field. We would like to pay attention 
to some of them. In Ref. \cite{Bar} the structures of field 
configurations were analyzed in full detail for a nonlinearity in NSE, 
which describes repulsion (in the absence of CS interaction) and takes 
into account the contribution of the non-zero mean value of  
particle number density. Considering the initial problem  
the authors of Ref. \cite{Ber} concluded that at the 
most general initial condition the problems of EM for NSE with 
CS gauge fields correspond to the collapse regime. However, 
neither the spatial structure of the collapsing 
mode, nor the critical power for it (i.e. the number of particles in 
the mode) have been analyzed in this paper. The problem of exact 
integration of the model under consideration was analyzed in 
Ref. \cite{Pas}. The main result obtained is that the system 
cannot be integrated exactly excluding the following two 
cases: self-dual limit \cite{Jac}, and the situation when the 
$2+1 D$ equations can be reduced to the $1+1 D$ equations. 
An additional but still very important  
result of that paper was that the solitons of the 
gauged nonlinear Schr\"odinger equation (GNSE) have movable 
singularities on some curves in the two-dimensional plane.  
Detailed study of topological defects in 
low-dimensional systems have always been a key point for 
understanding dynamics of field distributions. The problem of 
so-called semi-local topological defects in the 
Chern-Simons-Higgs model was considered in a recent paper 
\cite{Fue}. 

The distribution of complex-valued functions $\Psi (x,y,t)$ 
are defined on the manifold ${\cal M}$\/ 
which is multiconnected in spatial $2D$ systems. 
Therefore the fundamental homotopy group 
$\pi_{1}({\cal M})$\/ determining analytical properties of 
function $\Psi $ coincides with the braid group. There are actually 
several equivalent ways to reflect this fact in the theory. One of 
them is the Lagrangian approach that includes 
the effect of CS gauge fields into consideration. The CS term codes 
the existence and specific character of $2D$ point peculiarities 
contained in the Aharonov-Bohm gauge potentials within the long-wave 
description. One usually speaks of the long-range interaction 
represented by means of the CS gauge field as a statistical 
interaction between different field configurations. 
The different representations induce the different forms of field 
distribution. There is the so-called anyon representation 
\cite{Wil,Pro}, when 
the gauge field in the explicit form is excluded from the Hamiltonian of 
the model thus providing the representation of "non-interacting" 
(by means of the gauge field) configurations of field 
$\Psi (x,y,t)$. However, in this case the gauge field is 
proved to be included into the phase of function $\Psi 
(x,y,t)$ that contains a cut in the complex plane, which 
provides multivaluedness of the function. 
The cut describes a string attached to the point defect. 
Thus, the so-called 
non-local topological defect separates sheets in the multi-sheet 
coverage of the $2D$ basic space. 
It is well known 
\cite{Wil,Pro} that the representation with explicit presence of gauge 
fields in the Lagrangian has a different form  
depending on the parity of the permutation group representation. 
The approach, based on the representation with fractional statistics of field 
configurations, is identical to the dynamic approach, when we 
are interested in the influence of a "statistical" gauge field.

The existence of a gauge 
interaction has an exclusively topological character and is not 
associated with the quantum theory. This interaction, as a rule, 
was not taken into account when studying the classical dynamics of 
nonlinear models with a complex field in spatial $2D$ systems.  
Topological peculiarities certainly put additional restrictions 
to the quantization procedure in these systems \cite{Pro}. 
The role of the CS gauge interaction in this case is to take into 
account {\em the vortex part of phase dynamics}\/, which was not usually 
considered in classical systems when the $2+1 D$ NSE model was used.

The purpose of this paper is the study of EM in the $2+1 D$ GNSE 
model. The main attention is given to the investigation of the 
structure of collapsing distribution of the fields. Specifically 
we find, by means of numerical integration of the EM, 
the dependence of the critical 
power and efficient width of the zero-energy mode on coefficient 
$k$ before the CS term. The limit $k \to \infty$, when 
interaction with the gauge field is negligible, may be used as a 
test. In this case the known values of the power and the width 
are restored.

If the phase of the field $\Psi (x,y,t)$ describes the longitudinal 
part in the gauge potential {\em completely}\/, evolution of 
field configurations is determined {\em only by temporal 
dependence of the gauge field}\/. We show that in this case 
the equations for the gauge field coincide with the EM of an ideal 
fluid. The effects of the manifestation of gauge field in classic 
systems with nontrivial topology, including the swimming motion 
at low Reynolds number within the $2+1 D$ hydrodynamics are well 
known \cite{Sha}. A new feature is the fact that the basis for 
the 2D turbulence basing from the Euler equation in this case is 
chaotic dynamics of the CS gauge field. In this sense, the GNSE 
is a useful hydrodynamical tool \cite{Car}.  

One can see the following link between the dynamics of the CS 
fields and the problem of $2D$ turbulence. It is well known that 
CS action with appropriate boundary conditions is a way to classify 
conformal field theories \cite{Moo}. 
The tools of the conformal field theory, in its turn, may be 
used \cite{Pol} to study the $2D$ turbulence. We show that within 
the model under consideration the connection between the dynamics of CS 
fields and $2D$ turbulence problem may be stated beyond the  
application of the conformal field theory. This dependence can 
be represented considering the evolution of closed 
current lines (loops) with the stochastization of lines near 
the points of link of the loops. The effect of contour links in 
terms of this paper reflects the effect of braiding world lines 
of the Aharonov-Bohm point singularities with formation of 
knots and links after projection of world lines onto the $2D$ 
space. Stochastization near the contour link points within 
formulation of $2D$ hydrodynamics of an ideal fluid in terms of 
contour variables \cite{Zab,Mig} was discovered in Ref. 
\cite{Agi}. Such a stochastic behavior has a universal 
character. It is based on existence of the braid group and is 
closely connected with the arbitrary character of localization 
along the "time" axis of the point of world lines interlacing. 
Because of this the index $k$ of the linking number proves to be 
a hidden parameter, which is not included into the Euler equations 
explicitly.

The paper is organized as follows. The second section contains 
the EM for the CS gauge field and the GNSE for two different Ansatz 
corresponding to the goals of this paper. The third section 
is devoted to the numerical analysis of the problem. 
In conclusion, the results and open questions are discussed.  

\subsubsection*{2. Equations of motion}

We consider a system with a Lagrangian density
\begin{equation}
{\cal L} = {\frac {k} {2}}\varepsilon^{\alpha \beta \gamma}
A_{\alpha}\partial_{\beta}A_{\gamma} + i\Psi^{\ast}(\partial_{t}+
iA_{0})\Psi - 
{\frac {1} {2}}\left |(\nabla - i{\bf A})\Psi \right|^{2} + 
{\frac {g} {2}}\left|\Psi \right|^{4} \, .
\end{equation} 
The equations of motion (EM) have the form
\begin{equation}
i\partial_{t}\Psi = -{\frac {1} {2}}(\nabla - i{\bf A})^{2}\Psi 
+ A_{0}\Psi - g|\Psi |^{2}\Psi \, , 
\end{equation} 
\begin{equation}
[\nabla \times {\bf A}]_{\perp}=-{\frac {1} {k}}|\Psi|^{2} \, ,
\end{equation} 
\begin{equation}
\partial_{t}A_{i} + \partial_{i} A_{0} = -{\frac {1} {k}}
\varepsilon_{ij}j_{j} \, .
\end{equation} 
Here $g$ is the coupling constant and ${\bf j} = Im \Psi^{\ast}
(\nabla - i{\bf A})\Psi $ is the current density. 
Hamiltonian for Eq. (1) is 
\begin{equation}
H = {\frac {1} {2}}\int d^{2}{\bf r}
\left (|(\nabla - i{\bf A})\Psi |^{2} - g|\Psi |^{4}\right )\, , 
\end{equation} 
where the potential $A_{\mu}$ which is the auxiliary variable 
is expressed in terms of $|\Psi|^{2}$ in the following way 
\begin{equation}
{\bf A}({\bf r},t)={\frac {1} {k}}\int d^{2}{\bf r}^{\prime}
{\bf G}({\bf r}-{\bf r}^{\prime})|\Psi |^{2}({\bf r}^{\prime},t) \, , 
\end{equation} 
\begin{equation}
A_{0}({\bf r},t)={\frac {1} {k}}\int d^{2}{\bf r}^{\prime}
{\bf G}({\bf r}-{\bf r}^{\prime}){\bf j}
({\bf r}^{\prime},t) \, , 
\end{equation} 
where the Green function ${\bf G}({\bf r})$     
\begin{equation}
G_{i}({\bf r}) = {\frac {1} {2\pi}}
{\frac {\varepsilon_{ij}x_{j}} {r^{2}}}
\end{equation} 
satisfies the equation    
\begin{equation}
\nabla \times {\bf G}({\bf r})=-\delta^{2}({\bf r})
\end{equation} 
thus $A_{\mu}$ is the solution of Eqs. (3) and (4). 
Since in Hamiltonian formulation the potentials are 
unambiguously presented by Eqs. (6) and (7), 
the gauge freedom  
\begin{equation}
A_{\mu} \to A_{\mu}-\partial_{\mu}\varphi \, ,
\end{equation} 
\begin{equation}
\Psi \to e^{i\varphi}\,\Psi 
\end{equation} 
is fixed. This is achieved by the Coulomb gauge
$\nabla \cdot {\bf A}=0$, with boundary conditions                            
\begin{equation}
\lim_{r\to \infty}r^{2}A_{i}({\bf r},t)={\frac {1} {2\pi k}}
\varepsilon_{ij}x_{j}N \, ,
\end{equation} 
\begin{equation}
\lim_{r\to \infty}A_{0}({\bf r},t)=0 \, .
\end{equation} 
The choice of the boundary condition (12) is associated with 
the necessity to satisfy the integral representation of 
Gauss law (3) of CS dynamics 
\begin{equation}
\Phi=\int d^{2}{\bf r}\,[\nabla \times {\bf A}]_{\perp}
=-{\frac {1} {k}}\int d^{2}{\bf r}|\Psi|^{2} \, .
\end{equation} 
Here the magnetic flux $\Phi$ and the number of particles 
\begin{equation}
N=\int d^{2}{\bf r}|\Psi|^{2}
\end{equation} 
are conserved giving the global constrain 
$\Phi=-{\frac {1} {k}}N$. In the result of Eqs. (2)-(4), 
there exists naturally the continuity equation
\begin{equation}
\partial_{t}|\Psi|^{2}+\nabla \cdot {\bf j}=0 \, ,
\end{equation} 

Let us use dimensionless fields and coordinates 
obtained by the following substitutions 
\begin{equation} 
\Psi=|k|^{3/2}\rho e^{i\varphi}, \, \, \, \, \, \, 
A_{0}=-{\frac {k^{2}} {2}}w - \partial_{t}\varphi , \, \, \, \, 
\, \, 
A_{x}=-k\,u+\partial_{x}\varphi , \, \, \, \, \, \, 
A_{y}=-k\,v+\partial_{y}\varphi , \, \, \, \, \, \,  
\end{equation} 
\begin{equation}
t \to {\frac {-2} {k|k|}}t \, , \, \, \, \, \, \, 
x \to {\frac {x} {|k|}}\, , \, \, \, \, \, \, 
y \to {\frac {y} {|k|}}\, . 
\end{equation} 

The EM and the continuity equation, expressed by the 
new real functions $\rho \equiv \rho (x,y,t)$, 
$u\equiv u(x,y,t) $, $v\equiv v(x,y,t)$, 
$w\equiv w(x,y,t)$ have the form 
\begin{equation}
\rho_{xx}+\rho_{yy}=-2C\rho^{3}-\rho w+\rho(u^{2}+v^{2}) \, ,
\end{equation} 
\begin{equation}
u_{y}-v_{x}=-\rho^{2}\, ,
\end{equation} 
\begin{equation}
u_{t}-w_{x}=-2v\rho^{2} \, , 
\end{equation} 
\begin{equation}
v_{t}-w_{y}=2u\rho^{2} \, , 
\end{equation} 
\begin{equation}
\rho^{2}_{\, \, t}=2\left ((u\rho^{2})_{x}+
(v\rho^{2})_{y}\right )
\end{equation} 
with the parameter $C=g|k|$ and notations $u_{t}=\partial_{t}u$ etc.  

In the case of usual NSE 
\begin{equation}
i\partial_{t}\Psi = -\nabla^{2}\Psi - |\Psi|^{2}\Psi
\end{equation} 
substituting  $\Psi =\rho e^{-i\varphi (x,y,t)}$ we have
\begin{equation}
\rho_{xx}+\rho_{yy}=-\rho^{3}-\rho \varphi_{t} + 
\rho \left ((\varphi_{x})^{2}+(\varphi_{y})^{2}\right ) \, , 
\end{equation} 
\begin{equation}
\rho^{2}_{\, \, t}=2\left ((\varphi_{x}\rho^{2})_{x}+
(\varphi_{y}\rho^{2})_{y}\right ) \, .
\end{equation} 

Comparing Eqs. (25), (26) with Eqs. (19) and (23), we pay 
attention to the following distinctions. First, due to 
gauge invariance, there are no derivatives of phase 
$\varphi$ in Eq. (19) which exist in (25). Their role is played 
by gauge potentials. Therefore the evolution of the field  $\rho(x,y,t)$ 
is defined by the time derivatives of the functions $u(x,y,t)$ and 
$v(x,y,t)$ in Eqs. (20), (21). 
Unlike Eq. (25) the fields  $u$ and $v$  are responsible 
for {\em the transverse}\/ dynamics of the phase of the field $\Psi$. 
{\em Longitudinal}\/ dynamics of the phase is described 
by the zero component $w(x,y,t)$ of the gauge potential 
which takes the place of the function $\varphi_{t}$ in Eq. (19).  
The function $w(x,y,t)$ plays the role of a Lagrange multiplier 
permitting to take into account the restriction (20) 
of the Gauss law $\Phi=-{\frac {1} {k}}N$ locally. 

Second, the continuity equation (23) can be obtained excluding the 
function $w$ from Eqs. (20) and (21) with the aid of Eq. (20). 
This remark is associated with the following problem. Let us assume 
in the Coulomb gauge 
$\nabla \cdot {\bf A}=-u_{x}-v_{y}+\Delta \varphi = 0$ that the 
phase $\varphi $ satisfies the equation $\Delta \varphi$=0.  
Then the solution to the equation $u_{x}+v_{y}=0$ may be expressed 
by means of the function  $a(x,y,t)$ in the following way:  
\begin{equation}
u = a_{y}\, , \, \, \, \, \, \, \, \, \, v = -a_{x}\, .
\end{equation} 
In this case after replacing $t$ by $-2t$ 
Eqs. (20) and (23) have the form
\begin{equation}
a_{xx} + a_{yy} = -\rho^{2} \, ,
\end{equation} 
\begin{equation}
\rho^{2}_{\, \, t}+u\rho^{2}_{\, \, x}+v\rho^{2}_{\, \, y}
=0 \, .  
\end{equation} 

The set of Eqs. (28) and (29) represents the "vorticity" form of 
Navier-Stokes equations (Euler equations) for 
two-dimensional flows of ideal incompressible fluid where the 
function $a(x,y,t)$ has the sense of the stream function. Note 
that hydrodynamic analogies have already been used for the 
solution of $1+1 D$ NSE problem \cite{Sch,Car}. 
However, the exact proof that 
dynamics of CS gauge field in the frames of GNSE model 
(in the particular case of Coulomb gauge with 
$\Delta \varphi = 0$) is equivalent to the two-dimensional EM of 
ideal incompressible fluid is given in this paper for the 
first time. The remarkable fact is that there is a close 
analogy between the states with the constant flux in the 
turbulence and CS anomaly \cite{Pol} exposed by Eq. (28). 

It is useful to compare the gauge invariance of the model and 
the used Coulomb gauge at $\Delta \varphi = 0$ with canonical 
transformations and with area-preserving transformations. The 
infinitesimal area-preserving diffeomorphism which acts in the 
frames of CS theory has the form
\begin{equation}
\xi_{i} \to \xi_{i}+\chi_{i} \, , \, \, \, \, \, \, \, \, 
\partial_{i}\chi_{i}=0 \, ,
\end{equation} 
where $\xi_{i}=(x, y)$, $\chi_{i}=(A_{1}, A_{2})$. The 
general solution of the equation $\partial_{i}\chi_{i}=0$ is the 
sum of two terms \begin{equation}
\chi_{i}=\varepsilon_{ij}\partial_{j}a(\xi) + 
\sum\limits_{k=1}^{b_{1}}c_{k}\chi_{i}^{k} \, ,
\end{equation} 
where the second term describes the finite number
(given by the first Betti number  $b_{1}$) of harmonic forms 
on the two-dimensional phase space $(A_{x}, A_{y})$ of CS theory. 
Diffeomorphisms which resulted from the first term in Eq. (31) 
are nothing but canonical transformations \cite{Kog}. In the case  
of phase space which is actually a torus there are also two global   
translations $p_{k}=-i\partial_{k}$ (for torus the first 
Betti number $b_{1}=2$). Just in this case the phase $\varphi (x,y,t)$, 
satisfying the equation $\Delta \varphi = 0$, 
is the linear function  $\varphi=ax+by$. From the viewpoint of 
NSE this corresponds to the constant direction of ray 
propagation assigned by the vector ${\bf n} \sim (a,b)$. 
In the general case of the phase space with arbitrary 
topology it is invalid and the phase $\varphi (x,y,t)$ not 
satisfying the equation  $\Delta \varphi = 0$ gives 
rise to the "additional" longitudinal contribution to the 
potentials $u(x,y,t)$ and $v(x,y,t)$. 

Let us consider for example the case when the Ansatz for 
the field $\Psi (x,y,t)$ corresponds to the generalized lens 
transformation \cite{Ras,Ber} 
\begin{equation} \Psi ({\bf 
r},t)={\frac {\Phi ({\gir{$\zeta$}},\tau )} {g(\tau )} } \exp 
\left(-ib(\tau){\gir{$\zeta$}}^{2}/2 + 
i\lambda \tau \right) \; .
\end{equation} 
Here $\gir{$\zeta$} = {\bf r}/g(\tau)$, $\tau 
=\int\limits_{0}^{t}du \left[f(u)\right]^{-2}$, 
$b(\tau )=-f_{t}f=-g_{\tau}g$.  
The gauge potentials transform \cite{Jac}, at such 
substitution, are as follows
\begin{equation} 
{\bf A}({\bf r},t) 
\to \left[g(\tau) \right]^{-1} {\bf A}({\gir{$\zeta$}},\tau) \, 
, 
\end{equation} 
\begin{equation} A_{0}({\bf r},t) \to 
\left[g(\tau) \right]^{-2} 
\left[A_{0}({\gir{$\zeta$}},\tau)-b(\tau){\gir{$\zeta$}} 
{\bf A}({\gir{$\zeta$}},\tau)\right] 
\end{equation} 
the relations (6), (7), being preserved, where the function 
$\rho=|\Phi|$. After these transformations Eq. (2) 
changes its form 
\begin{equation} 
i\partial_{\tau}\Phi + (\beta {\gir{$\zeta$}}^{2}-\lambda )\Phi=
-{\frac {1} {2}}(\nabla - i{\bf A})^{2}\Phi + A_{0}\Phi - 
g|\Phi|^{2}\Phi \, ,
\end{equation} 
because the function $\beta 
(\tau)=(b^{2}+b_{\tau})/2=-f^{3}f_{tt}/2$ in the case of 
$\varphi (x,y,t) \sim b(x^{2}+y^{2})$ and $b(t)\neq t_{0}-t$ 
does not equal zero. If we are interested in collapsing 
solutions with  \cite{Fra,Ber} á $f^{2}(t) \sim (t_{0}-t)/\ln 
\left[\ln (t_{0}-t)\right]$, the structure of the self-similar 
nonlinear core \cite{Ber} of the solution is described by the 
solutions of the following equation 
\begin{equation} 
-\lambda 
\Phi = -{\frac {1} {2}}(\nabla - i{\bf A})^{2}\Phi + A_{0}\Phi - 
g|\Phi|^{2}\Phi \, .
\end{equation} 

In the next section, by numerical calculation, we find the 
zero-energy localized ground state of GNSE (36).  
We show the dependence of its effective width on the 
parameter $C=g|k|$ as well as the form of the functions $u, v, w$.

\subsubsection*{3. Solution structure}

For the numerical analysis of the solutions of Eq. (36) 
we use the method of the stabilizing multiplier \cite{Pet}.  
The iteration approach for Eq. (36), which differs from 
Eq. (19) by the additional term $-\lambda \Phi$ in the LHS 
has the form   
\begin{equation} 
\Phi_{n+1}=M_{n}F^{-1}\left(
G(p)F\left(-2C\Phi_{n}^{3}+j\Phi_{n}(u^{2}
+v^{2}-w)_{n}\right)
\right)  \, ,
\end{equation} 
\begin{equation} 
M_{n}=\left({\frac {\int d^{2}p (F\Phi_{n})^{2}} 
{\int d^{2}p 
G(p)
F\Phi_{n}
F(-2C\Phi_{n}^{3}+j\Phi_{n}(u^{2}+v^{2}-w)_{n})}}\right)^{\alpha}\,. 
\end{equation} 
Here $F$ $(F^{-1})$ are the operators of direct (inverse) 
Fourier transform, $G(p)=-\left(p^{2}+\lambda\right)^{-1}$. 
The multiplier $j=1$ or $j=0$ if we take into account the 
nonlinear contribution of gauge fields in Eq.(36) or neglect it. 
In the case $j=0$ the usual normalization in 
NSE corresponds to $C=1/2$. Without the restriction of 
generality we shall suppose below that $\lambda =1$. 

We should choose the exponent $\alpha$ in the stabilizing 
multiplier $M_{n}$ comparing the degrees of homogeneity of terms in LHS 
and RHS of Eq. (36) proceeding from the requirements that $M_{n} 
\to 1$ at $n \to \infty$. Without the term $\Phi 
(u^{2}+v^{2}-w)$ the exponent $\alpha$ equals $3/2$. New 
features of our problem are that the nonlinearity in Eq. (36) 
has the polynomial character of the type $-2C\Phi^{3}+b\Phi^{5}$ 
because both of the terms $\Phi w$ and $\Phi (u^{2}+v^{2})$ are 
proportional to $\Phi^{5}$. Therefore, for the convergence of the 
iteration approach $\alpha$ should belong to the range $5/4 \leq 
\alpha \leq 3/2$. In the simulation of the present paper we have 
used the value $\alpha =3/2$ which gives rapidly the value 
$M_{n}=1$ of the stabilizing multiplier. We have used the 
distribution of the form $\Phi(\zeta_{x},\zeta_{y})=
\left(\sqrt{\gamma \delta}/\pi \right)\exp (-\gamma 
{\zeta_{x}}^{2}-\delta {\zeta_{y}}^{2})$ as the initial field 
configurations with arbitrary constants $\gamma$ and $\delta$. 
In our calculations we have obtained rapidly the isotropic solutions. 

To regularize the integrals (6) and (7) which diverge 
logarithmically in coincident points at numerical 
calculations of gauge potentials $u$ and $v$ we substituted 
$r^{2} \to r^{2}+\varepsilon^{2}$ in the expression for Green 
function (8). In the momentum space this corresponds 
to the substitution of $d^{3}p\,f(\varepsilon p)$
with $f(\varepsilon p)=\int \limits_{0}^{\infty}dm \exp (-\sqrt 
{m^{2}+\varepsilon^{2}p^{2}})$ for $d^{3}p$. The factor 
$f(\varepsilon p)$ at $\varepsilon p \gg 1$ decreases 
exponentially cutting off all momentum integrals. However, the 
infrared region remains the same, because at 
$\varepsilon p \ll 1$ $f(\varepsilon p)=1$. 
In our calculations the cut-off radius $\varepsilon $ which has the 
sense of the thickness of vortex core, was equal to $10^{-2}$. 

Simulation was performed on the square lattice with 
linear sizes $L_{x}=L_{y}=12$. The maximum number of 
lattice sites was limited by the value $n=n_{x}n_{y}=128\cdot 128$. 
To test our approach we used the solution of EM (36) 
with $A_{\mu}=0$ $(j=0)$ and with $C=1/2$ which gives the 
well-known value $N=11.703$, as well as the solution of a 
self-dual equation $\Delta \ln \rho = -\rho^{2}$ \cite{Jac}, 
when $w=-\rho^{2}$, $u=\partial_{y}\ln \rho$, 
$v=-\partial_{x}\ln \rho$.  

The Figures 1-3 show the configurations of the fields  $\rho $, 
$u$ and $w$ for the specific value of the parameter $C=4$. 
We may obtain the form of the function $v(\zeta_{x},\zeta_{y})$ 
using the relation $v(\zeta_{x},\zeta_{y})=-u(\zeta_{y},\zeta_{x})$.  

Using the obtained function $\rho$ we computed the 
dependencies of the critical power $N$ (the particle number) and 
the effective width $<R^{2}> =N^{-1}\int d^{2}\zeta\, 
{\gir{$\zeta$}}^{2}\rho ({\gir{$\zeta$}})$ on the parameter $C$. 
The results of calculations are given in the Table 1.
\vspace{1.5cm}

Table 1.\\ 

\begin{tabular}{|c||c|c|c|}                      \hline 
$j$ & C          & N          & $<R^{2}>$    \\ \hline \hline 
$0$ & $0.5$      & 11.703     & 1.2607       \\ \hline 
$1$ & $3$        & 2.9216     & 1.2464       \\ \hline 
$1$ & $5$        & 1.2825     & 1.2579       \\ \hline 
$1$ & $10$       & 0.5973     & 1.2600       \\ \hline 
$1$ & $100$      & 5.8528$\cdot 10^{-2}$  & 1.26066 \\ \hline 
$1$ & $1000$     & 5.8516$\cdot 10^{-3}$  & 1.26066  \\ \hline 
\end{tabular} 
\vspace{0.25cm}

\subsubsection*{4. Conclusion}

In this paper we have studied systematically the influence of 
the CS gauge field, reflecting the specific feature of the 
dimensionality of our problem, on field configurations in the 
GNSE model. Here we summarize some new results.     

It is seen from Eqs. (15)-(19) that if we neglect CS 
gauge fields ($j=0$ in Eq. (37)) the dependence of the 
particle number $N$ on the parameter $C=g|k|$ has the form $N=N_{0}/C$. 
This dependence is shown in Fig.4 by the dotted line. It follows 
from the results shown in the first line of Table 1 that 
$N_{0}=5.585$. The contribution of CS gauge fields (j=1 in the 
Table 1) leads to the sharp decrease of values of $N$. 
In particular $N_{j=1}(3)/N_{j=0}(0.5) \approx 0.25$. 
The effective width $<R^{2}>$ changes slightly. 
The calculated dependence $N(C)$ at $C \geq 3$ is given in Fig.4. 

As it should be expected at a fixed value of the parameter $C$ 
in the region $C \geq 3$, $N_{j=1}(C)$ is always greater than 
$N_{j=0}(C)$, because the CS gauge fields describe effective repulsion.
In the range $1 < C <3$ we could not perform calculations in the 
framework of the method used, due to the breaking of convergence of the 
iteration method (37), (38). We cannot explain this phenomenon using 
the theory presented in this paper. 
Note however, that the values $C=1$ and $C=2$ at $g=1$ correspond to 
the discrete values $k=1$ and $k=2$ which 
describe the field configurations in the self-dual limit 
$k=1$ \cite{Jac} and the one-half fermion limit $k=2$, which 
qualitatively differ from our case. The classic limit of the 
considered theory corresponds to the case $k \to \infty$, when the 
gauge field splits off from the field $\Psi (x,y,t)$.  For Eq. 
(19) and for the value $N(C)$ the limit $C \gg 1$ denotes that 
$N(C) \to 0$. 

Strong Langmuir turbulence in plasmas is usually described by 
the solutions of the NSE (Eq. 24). It is assumed that a cascade 
of the randomly distributed self-similar collapsing fields is generated. 
In this paper we show that the specific features of spatial two-dimensional 
systems may lead to the traditional picture of turbulence associated 
with Euler equations. 
However, for the hydrodynamical mechanism of turbulence (HMT) 
to be involved, it is necessary the existence of a linear profile 
of the phase $\varphi (x,y)$ ($=ax + by$) in each mode. This implies 
that the nonlinear contributions (by $x$ and $y$) in the temporal 
evolution of the phase are small. 

One of the media in which the HMT can play a role is an optical 
medium with random inhomogeneous guiding surfaces. 
Reflecting from the surfaces, wave fronts acquire random directions of 
propagation. For media with weak Kerr nonlinearity, the nonlinear 
phase disturbance from adjacent points will not be important. 
Application of the HMT model suggested above requires separate 
consideration and will be presented elsewhere.   

In conclusion, we studied numerically the structure of the 
collapsing mode in GNSE model, observed the effect of 
strong reduction of the critical power $N$ in spatial 
two-dimensional systems as compared to the traditional values,  
and showed that in the case of appropriate boundary conditions 
the phenomenon of collapse inhibits the development of turbulence 
according to the hydrodynamic scenario.

\subsubsection*{Acknowledgments}

We would like to thank G.M. Fraiman, E.A. Kuznetsov, 
L.A. Litvak, V.A. Mironov, A.M. Sergeev, S.N. Vlasov and A.D. 
Yunakovsky for numerous stimulating discussions and useful 
comments. One of the authors (A. P.) would like to thank 
the ICTP for hospitality throughout a visit during 
which this manuscript was completed. Numerical simulations 
were made using the workstation 
offered by the European Union -DG III/ESPRIT - 
Project CTIAC 21042. This work was supported in part by the 
Russian Basic Research Foundation under Grant No. 95-02-05620, 
by the High School Committee of the Russian Federation under 
Grant No.  95-0-7.4-173, as well as by the ISI Foundation and 
EUINTAS Network 1010-CT 930055.

\newpage
\vspace{4cm}

\begin{center}
{\bf Figure captions}
\end{center}

\vspace{0.5cm}

{FIG. 1. 
\parbox[t]{110mm}{
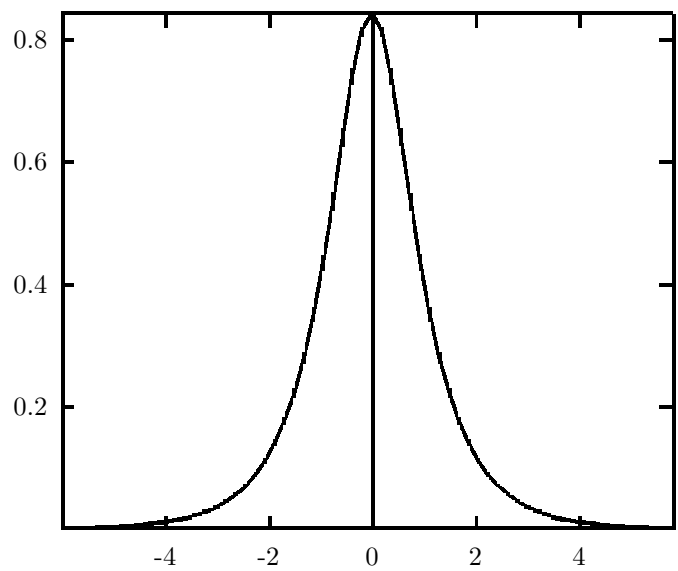
Plot of the function $\rho (\zeta_{x},0)=\rho 
(0,\zeta_{y})$.}}\\  

{FIG. 2. 
\parbox[t]{100mm}{
Plot of the function $u(\zeta_{x},0)$.}} \\  

{FIG. 3. 
\parbox[t]{100mm}{
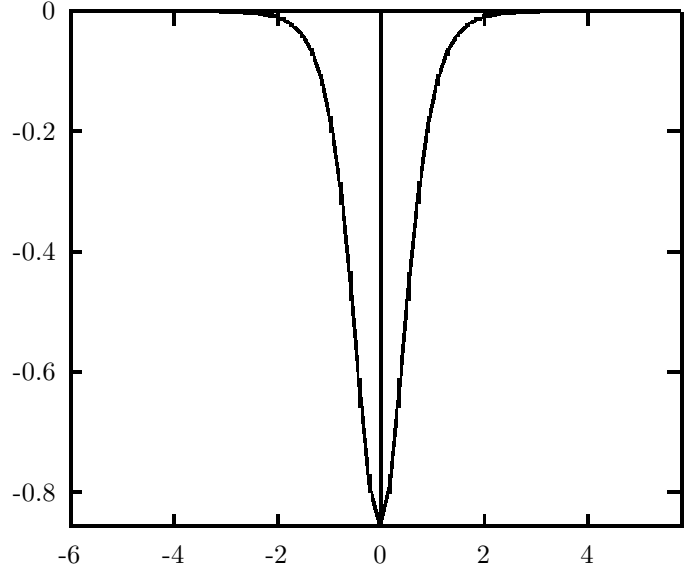
Plot of $w(\zeta_{x},0)=w(0,\zeta_{y})$.}} \\  

{Fig. 4. 
\parbox[t]{120mm}{
Number of particles $N$ as a function of the parameter $C$ 
without taking into account the gauge field (dotted line)
and with the gauge field (solid line). The point denotes the 
value $N(0.5)=11.703$.}} 





\end{document}